\documentclass{article}

\usepackage{microtype}
\usepackage{graphicx}
\usepackage{caption}
\usepackage{booktabs} 
\usepackage{listings}
\usepackage{xcolor} 
\usepackage{float} 
\usepackage{tabularx}
\usepackage{tikz}
\usetikzlibrary{arrows.meta, positioning, decorations.pathreplacing, shadows}
\usepackage{placeins}
\usepackage{algorithm}
\usepackage{algorithmic}

\usepackage{tcolorbox}   
\tcbset{
    mypythonbox/.style={
        listing engine=listings,
        listing only,
        breakable,
        enhanced,
        colback=gray!5,
        colframe=black,
        boxrule=0.5pt,
        arc=2mm,
        outer arc=2mm,
        left=2mm,
        right=2mm,
        top=1mm,
        bottom=1mm,
        title=#1,
    }
}

\usepackage{hyperref}

\usepackage[preprint]{icml2026}

\usepackage{amsmath}
\usepackage{amssymb}
\usepackage{mathtools}
\usepackage{amsthm}

\usepackage[capitalize,noabbrev]{cleveref}

\theoremstyle{plain}

\theoremstyle{definition}

\theoremstyle{remark}

\usepackage[textsize=tiny]{todonotes}

\icmltitlerunning{Hardware-Aware Reformulation of CNNs for Efficient Execution on Specialized AI Hardware}

\begin{document}

\twocolumn[
	\icmltitle{Hardware-Aware Reformulation of CNNs for Efficient Execution on Specialized AI Hardware: A Case Study on NVIDIA Tensor Cores}



  \icmlsetsymbol{equal}{*}

  \begin{icmlauthorlist}
	  \icmlauthor{Ganesh Bikshandi}{yyy}
  \end{icmlauthorlist}
  \icmlaffiliation{yyy}{vbganesh79@gmail.com}
  \icmlcorrespondingauthor{Ganesh Bikshandi}{vbganesh79@gmail.com}

  \icmlkeywords{Machine Learning, ICML}

  \vskip 0.3in
]
\printAffiliationsAndNotice{}  

\begin{abstract}
	Convolutional Neural Networks (CNNs) are central to modern AI, but their performance is often limited by hardware constraints. NVIDIA Tensor Cores, for instance, require input channels to be multiples of 8 and sometimes 512 for efficient execution. {\em oneDNN} framework for CPU imposes such a requirement for the blocked format. Traditional approaches address such alignment issue using zero-padding, which can be inefficient. In this work, we present a first-step, hardware-aware reformulation of CNN computations using rewrite rules, restructuring the underlying math to satisfy hardware alignment entirely {\bf post-training} without modifying network weights. While our current implementation focuses on a single transformation for Tensor Cores, this approach is generalizable, laying the foundation to explore additional transformations for CPU and accelerators. This study represents an initial step toward {\em semantic tuning}, a systematic, hardware-aware optimization strategy for efficient deployment of CNN models on specialized AI hardware.
\end{abstract}

\section{Introduction}

Convolutional Neural Networks (CNNs) are a foundational building block of modern deep learning systems, underpinning applications ranging from computer vision to speech and scientific computing. At a high level, CNNs apply learnable filters to structured input tensors in order to extract hierarchical feature representations. For a standard convolution, the input tensor has shape $(N, C_{\text{in}}, H[, W, D])$, the filter weights have shape $(C_{\text{out}}, C_{\text{in}}, K_1[, K_2, \ldots, K_n])$, and the resulting output tensor has shape $(N, C_{\text{out}}, H_{\text{out}}[, W_{\text{out}}, D_{\text{out}])}$. This formulation uses the NCHW layout; alternate layout is NHWC.

While this mathematical formulation is hardware-agnostic, the efficient execution of CNNs on modern AI accelerators is strongly influenced by hardware-specific constraints. Specialized units such as NVIDIA Tensor Cores and analogous matrix-multiply engines in other accelerators impose alignment and tiling requirements on tensor dimensions, most notably that certain dimensions (e.g., channel counts) be multiples of fixed factors such as eight or a higher count like 512~\cite{nvidia_cnn}. Also, {\em oneDNN} framework~\cite{onednn2025} for CPU imposes such an alignment requirement for the blocked format. When these constraints are not satisfied, libraries either fall back to less efficient execution paths or require auxiliary modifications such as zero padding, which introduce redundant computation and memory overhead.

It is possible to address these constraints during network design or training, for example by manually increasing the channels or retraining models with $1\times1$ convolution. In contrast, this paper explores a complementary and largely underexplored direction: post-training reformulation of CNN computations. Rather than modifying the network architecture or retraining weights, we rewrite the underlying convolutional mathematics to produce an equivalent formulation that satisfies hardware alignment and channel count requirements by construction.

Specifically, we present an initial hardware-aware transformation that reshapes and reinterprets the convolutional computation—through {\bf width folding for input in NHWC format (alternatively height folding for NCHW format)} and {\bf structured filter expansion} — so that the resulting tensors conform to accelerator alignment constraints while preserving exact numerical equivalence to the original CNN. This transformation eliminates the need for zero padding and does not alter the learned parameters or outputs of the network. Although the paper focuses on a single transformation motivated by NVIDIA Tensor Core constraints, the broader goal is to establish a foundation for a family of such rewrite rules that can be systematically derived and applied across different accelerator architectures.

Viewed through this lens, CNN execution can be treated as a compilation problem, where mathematically equivalent reformulations are selected to best match the capabilities and constraints of the target hardware. This work represents a first step toward such a hardware-aware compilation framework for CNNs, demonstrating that non-trivial accelerator constraints can be addressed purely through post-training mathematical rewrites.

\begin{table}[h!]
\centering
\small
\caption{First-layer channels in popular CNN architectures}
\label{tab:firstlayer}
\begin{tabularx}{\columnwidth}{|X|c|}
\hline
\textbf{Network} & \textbf{Input Channels} \\
\hline
AlexNet  & 3 \\
VGG16  & 3 \\
ResNet-18 & 3 \\
ResNet-50 & 3 \\
GoogLeNet & 3 \\
MobileNetV2 & 3 \\
\hline
\end{tabularx}
\end{table}

To motivate the need for hardware-aware filter alignment, we survey several widely used CNN architectures and their first-layer channel dimensions.  As shown in Table~\ref{tab:firstlayer}, most models processing RGB images have an input channel count of 3, which is not a multiple of 8, limiting the efficiency on NVIDIA Tensor Cores. Furthermore, many models have a first-layer filter count below 512, which is the recommended number mentioned in NVIDIA~\cite{nvidia_cnn}. In such cases, our method can also increase the channel dimensions, ensuring better alignment with hardware requirements while preserving network functionality.

The paper makes the following contributions:

The paper makes the following contributions:

\begin{itemize}
  \item Introduce \emph{semantic reformulation} as a hardware-aware optimization paradigm.
  \item Present \emph{width folding}, a semantics-preserving transformation that increases effective channel dimensionality without padding.
  \item Provide a formal correctness proof.
  \item Show how the transformation fits naturally as an MLIR compiler pass.
  \item Demonstrate generalization to GEMM via $1\times1$ convolutions.
  \item Report up to $3\times$ speedups over cuDNN/TensorRT on NVIDIA A100.
\end{itemize}

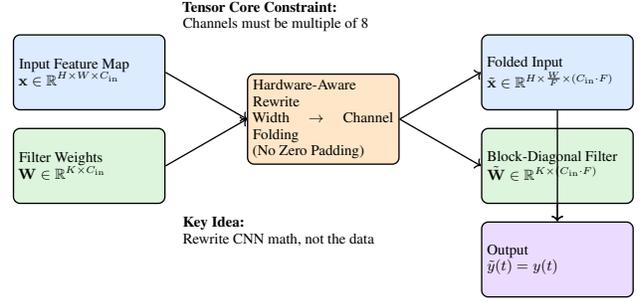
\begin{figure}[h!]
\centering
\resizebox{\columnwidth}{!}{%
\begin{tikzpicture}[
    box/.style={draw, thick, rounded corners, minimum width=3.2cm, minimum height=1.6cm, align=center},
    arrow/.style={->, thick},
    every node/.style={font=\small}
]

\definecolor{inputblue}{RGB}{220,235,255}
\definecolor{filtergreen}{RGB}{220,245,220}
\definecolor{rewriteorange}{RGB}{255,230,200}
\definecolor{outputpurple}{RGB}{235,220,255}

\node[box, fill=inputblue] (input1) at (0,2) {%
\parbox{3cm}{Input Feature Map\\
$\mathbf{x} \in \mathbb{R}^{H \times W \times C_{\rm in}}$}%
};

\node[box, fill=filtergreen] (filter1) at (0,0) {%
\parbox{3cm}{Filter Weights\\
$\mathbf{W} \in \mathbb{R}^{K \times C_{\rm in}}$}%
};

\node[box, fill=rewriteorange] (rewrite) at (5,1) {%
\parbox{3cm}{Hardware-Aware Rewrite\\
Width $\rightarrow$ Channel Folding\\
(No Zero Padding)}%
};

\node[box, fill=inputblue] (input2) at (10,2) {%
\parbox{3cm}{Folded Input\\
$\tilde{\mathbf{x}} \in \mathbb{R}^{H \times \frac{W}{F} \times (C_{\rm in} \cdot F)}$}%
};

\node[box, fill=filtergreen] (filter2) at (10,0) {%
\parbox{3cm}{Block-Diagonal Filter\\
$\tilde{\mathbf{W}} \in \mathbb{R}^{K \times (C_{\rm in} \cdot F)}$}%
};

\node[box, fill=outputpurple] (output) at (10,-2) {%
\parbox{3cm}{Output\\
$\tilde{y}(t) = y(t)$}%
};

\draw[arrow] (input1.east) -- (rewrite.west);
\draw[arrow] (filter1.east) -- (rewrite.west);

\draw[arrow] (rewrite.east) -- (input2.west);
\draw[arrow] (rewrite.east) -- (filter2.west);

\draw[arrow] (input2.south) -- (output.north);
\draw[arrow] (filter2.south) -- (output.north);

\node at (5,3.2) {%
\parbox{6cm}{\textbf{Tensor Core Constraint:}\\Channels must be multiple of 8}%
};

\node at (5,-1.4) {%
\parbox{6cm}{\textbf{Key Idea:}\\Rewrite CNN math, not the data}%
};

\end{tikzpicture}%
}
\caption{Hardware-aware reformulation of a CNN via width-to-channel folding.
The rewrite preserves exact semantics while satisfying Tensor Core alignment
constraints without zero padding or retraining.}
\label{fig:cnn_rewrite}
\end{figure}

\begin{figure}[t]
\centering
\resizebox{\columnwidth}{!}{%
\begin{tikzpicture}[font=\small,>=stealth]

\tikzset{
    tensor/.style={draw, thick, rounded corners=2pt, top color=#1!20, bottom color=#1!60, shading angle=45},
    depth/.style={draw, thick, fill=#1!40, opacity=0.85},
    kernel3d/.style={draw, thick, left color=gray!20, right color=gray!60},
    cell/.style={draw, thick, fill=gray!12},
    diag/.style={draw, thick, fill=red!65, drop shadow},
    arrow/.style={->, thick, color=blue!70}
}

\node[tensor=blue, minimum height=2.2cm, minimum width=4.2cm] (orig) at (0,0) {};
\node[above=4pt of orig] {\parbox{2.5cm}{\centering Input tensor}};
\node[below=4pt of orig] {\parbox{2.5cm}{\centering $H \times W,\; C_{in}=1$}};

\draw[arrow] (orig.east) to[out=0,in=180] ++(1.6,0)
    node[midway,above]{\parbox{2.5cm}{\centering Width folding ($F=8$)}};

\begin{scope}[shift={(6,0)}]
    \foreach \i/\c in {0/red,1/orange,2/yellow,3/green,4/cyan,5/blue,6/purple,7/magenta} {
        \node[depth=\c, minimum height=2.2cm, minimum width=0.6cm] at (\i*0.18,\i*0.18) {};
    }
    \node at (1.6,1.9) {\parbox{3cm}{\centering Folded input}};
    \node at (1.6,-1.5) {\parbox{3cm}{\centering $H \times (W/8)$}};
    \node at (1.6,-2.1) {\parbox{3cm}{\centering $C_{in}' = 8$}};
    \node[align=center] at (1.6,-2.8) {\scriptsize\parbox{3cm}{\centering Channels stacked in depth}};
\end{scope}

\begin{scope}[shift={(0,-4.8)}]
    \node[kernel3d, minimum width=0.45cm, minimum height=2.2cm] (filt) at (0,0) {};
    \node[above=4pt of filt] {\parbox{2.5cm}{\centering Original filter}};
    \node[below=4pt of filt] {\parbox{2.5cm}{\centering $K \times 1$}};
\end{scope}

\draw[arrow] (1.1,-4.8) to[out=0,in=180] (2.6,-4.8)
    node[midway,above]{\parbox{2.5cm}{\centering Channel expansion}};

\begin{scope}[shift={(5.2,-6.6)}, yscale=-1]
    \begin{scope}[opacity=0.15]
        \foreach \i in {0,...,7} {
            \foreach \j in {0,...,7} {
                \node[cell, fill=black] at (\j*0.5+0.08, \i*0.5+0.08) {};
            }
        }
    \end{scope}

    \foreach \i in {0,...,7} {
        \foreach \j in {0,...,7} {
            \ifnum\j=\i
                \node[cell,diag] at (\j*0.5,\i*0.5) {};
            \else
                \node[cell] at (\j*0.5,\i*0.5) {};
            \fi
        }
    }

    \node at (1.75,4.3) {\parbox{3cm}{\centering Expanded filter}};
    \node at (1.75,-1.1) {\parbox{3cm}{\centering Diagonal replication}};
    \node at (1.75,-1.7) {\parbox{3cm}{\centering $C_{in}' = 8$}};
\end{scope}

\end{tikzpicture}%
}
\caption{
Semantic-preserving CNN reformulation via width folding.
The input width is partitioned into $F=8$ interleaved slices and stacked along the channel dimension, increasing the effective number of input channels without altering spatial height.
The original $K \times 1$ convolution kernel is replicated along the main diagonal of the expanded filter matrix, ensuring independent convolution of each folded slice.
This post-training transformation preserves exact convolution semantics while aligning channel dimensions with accelerator constraints.
}
\label{fig:full-width-folding}
\end{figure}
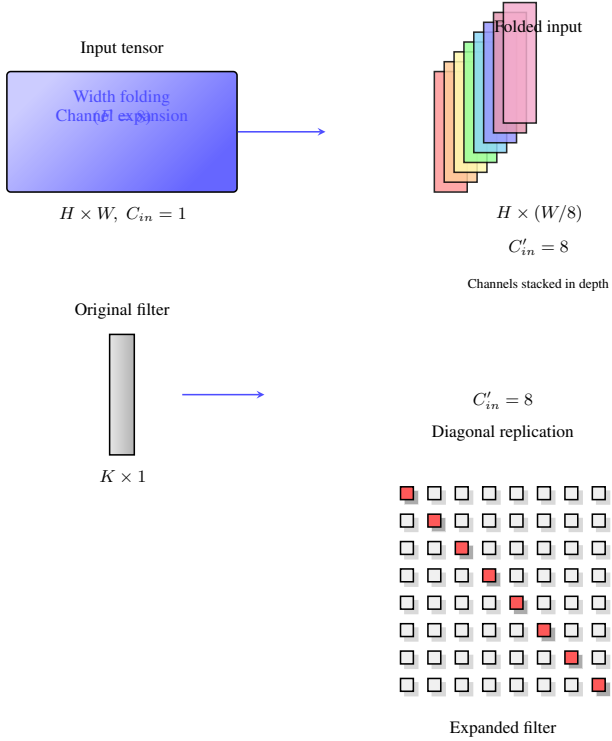


\begin{figure}[t]  
\centering
\small  
\resizebox{\columnwidth}{!}{%
\begin{minipage}{\columnwidth}
\begin{algorithm}[H]
\caption{Width-Folding Transformation for Convolution}
\label{alg:width-folding}
\begin{algorithmic}[1]
\REQUIRE
Input tensor $X \in \mathbb{R}^{B \times H \times W \times C_{\text{in}}}$, \\
Filter tensor $W_f \in \mathbb{R}^{K_H \times K_W \times C_{\text{in}} \times C_{\text{out}}}$, \\
Bias $b \in \mathbb{R}^{C_{\text{out}}}$, \\
Folding factor $F$

\ENSURE
Transformed tensors $(X_f, W_f', b')$ or fallback

\IF{$W \bmod F \neq 0$ \textbf{or} $C_{\text{in}} \neq 1$}
    \STATE \textbf{return} $(X, W_f, b)$ \COMMENT{Fallback: width not divisible or unsupported channels}
\ENDIF

\STATE Define $X_f \in \mathbb{R}^{B \times H \times (W/F) \times F}$

\FOR{$b\_idx = 0$ to $B-1$}
    \FOR{$h = 0$ to $H-1$}
        \FOR{$w' = 0$ to $(W/F)-1$}
            \FOR{$f = 0$ to $F-1$}
                \STATE $X_f[b\_idx,h,w',f] \gets X[b\_idx,h,F \cdot w' + f,0]$
            \ENDFOR
        \ENDFOR
    \ENDFOR
\ENDFOR

\STATE Define $W_f' \in \mathbb{R}^{K_H \times K_W \times F \times (F \cdot C_{\text{out}})}$, initialize to zero

\FOR{$f = 0$ to $F-1$}
    \STATE $W_f'[:,:,f,f\cdot C_{\text{out}}:(f+1)\cdot C_{\text{out}}] \gets W_f[:,:,0,:]$
\ENDFOR

\STATE Define $b' \in \mathbb{R}^{F \cdot C_{\text{out}}}$

\FOR{$f = 0$ to $F-1$}
    \STATE $b'[f\cdot C_{\text{out}}:(f+1)\cdot C_{\text{out}}] \gets b$
\ENDFOR

\STATE \textbf{return} $(X_f, W_f', b')$
\end{algorithmic}
\end{algorithm}
\end{minipage}%
}
\end{figure}

\section{Width Folding Transformation}

This section describes the width folding transformation, a post-training,
semantics-preserving reformulation that increases the effective channel
dimension to satisfy hardware alignment constraints while leaving the learned
model parameters unchanged.

Given an input tensor
\[
X \in \mathbb{R}^{H \times W \times C_{in}}, \quad C_{in}=1,
\]
and a 1-D convolution kernel
\[
W_f \in \mathbb{R}^{K \times 1},
\]
we fold the width dimension by a factor $F$ such that $W$ is divisible by $F$.
The transformation produces an equivalent convolution with
\[
C_{in}' = F, \quad W' = W / F,
\]
while preserving exact convolution semantics.

\subsection{Input Width Folding}

The width folding operation partitions the width dimension into $F$ interleaved
slices and stacks them along the channel dimension.
Formally, the transformed input tensor
\[
X' \in \mathbb{R}^{H \times (W/F) \times F}
\]
is defined as
\begin{equation}
X'(h, w', f) = X(h, Fw' + f),
\quad
f = 0,\ldots,F-1.
\label{eq:width_fold_def}
\end{equation}

This operation is a pure re-indexing and does not alter the numerical values of
the input tensor.

\subsection{Filter Construction}

Because convolution is performed only along the height dimension, each width
slice is convolved independently.
To preserve this behavior after folding, the original filter is replicated
across the expanded channel dimension without introducing cross-channel mixing.

Let the original filter be
\[
W_f \in \mathbb{R}^{K \times 1}.
\]
We construct a new filter tensor
\[
W_f' \in \mathbb{R}^{K \times F \times F}
\]
as a diagonal replication:
\begin{equation}
W_f'(k, f, f')
=
\begin{cases}
W_f(k), & f = f', \\
0, & f \neq f'.
\end{cases}
\label{eq:filter_construct}
\end{equation}

In implementation terms, this corresponds to allocating a zero-initialized
filter tensor and copying the original filter into the diagonal channel blocks.
Conceptually, each folded width slice receives an identical copy of the
original kernel.

\subsection{Bias Construction}

The original convolution bias
\[
b \in \mathbb{R}
\]
is shared across all folded slices.
Accordingly, the new bias vector is constructed by replication:
\begin{equation}
b'(f) = b,
\quad
f = 0,\ldots,F-1.
\label{eq:bias_construct}
\end{equation}

This ensures that each expanded channel applies the same bias as in the
original formulation.

\subsection{Resulting Convolution}

After width folding and filter construction, the transformed convolution
operates on
\[
X' \in \mathbb{R}^{H \times (W/F) \times F}
\]
using the filter $W_f'$ and bias $b'$.
As shown in Section~\ref{sec:correctness}, this convolution produces outputs
that are exactly equivalent to those of the original network, up to a
bijective re-indexing of the width dimension. The algorithm for computing
width folded convolution is presented in ~\ref{alg:width-folding}.

\section{Width Folding: Mathematical Perspective}

Width folding is a structural transformation of convolutional tensors that trades spatial width for additional channels. Given an input tensor 
$X \in \mathbb{R}^{B \times H \times W \times C_{\text{in}}}$ 
and a folding factor $F$, width folding produces 
$X_f \in \mathbb{R}^{B \times H \times (W/F) \times (C_{\text{in}} \cdot F)}$, via a linear isomorphism
\[
X_f[b,h,w',c'] = X[b,h,F \cdot w' + f, c], \quad c' = f \cdot C_{\text{in}} + c,
\]
preserving all input information. Conceptually, this is equivalent to reindexing the contraction indices in convolution, where the standard contraction
\[
Y[b,h,w,c_{\text{out}}] = \sum_{k_h,k_w,c_{\text{in}}} X[b,h+k_h,w+k_w,c_{\text{in}}]\, W[k_h,k_w,c_{\text{in}},c_{\text{out}}]
\]
becomes, after width folding,
\[
Y_f[b,h,w',c'_{\text{out}}] = \sum_{k_h,k_w,c'_{\text{in}}} X_f[b,h+k_h,w',c'_{\text{in}}]\, W_f'[k_h,k_w,c'_{\text{in}},c'_{\text{out}}].
\]

In linear algebra terms, folding flattens width into channels, allowing the convolution to be represented as a block-diagonal matrix multiplication, where each block corresponds to one slice along the folded width. This structure is naturally interpreted as a Kronecker product: the expanded kernel $W_f'$ can be viewed as the Kronecker product of the original kernel with an identity along the folded width dimension, which aligns well with hardware vectorization and parallelism.

From a category-theoretic viewpoint, tensors are objects in a monoidal category (e.g., $\mathbf{Vect}_\mathbb{R}$) and convolution is a morphism 
$X \otimes W \to Y$. Width folding is a natural isomorphism of the form
\[
B \otimes H \otimes W \otimes C_{\text{in}} \cong B \otimes H \otimes W' \otimes (C_{\text{in}} \otimes F),
\]
reassociating the tensor factors without changing the underlying morphism. Thus, folding is a structure-preserving transformation: the semantics of convolution remain identical, while the indexing structure changes to facilitate computation.

Overall, width folding unifies perspectives from tensor reshaping, contraction, block-diagonal / Kronecker structures, and categorical isomorphisms, providing a concise mathematical framework for this hardware-oriented optimization.

\section{Correctness of Width Folding Transformation}
\label{sec:correctness}

We consider a 1-D convolution performed exclusively along the height
dimension \(H\).
The width dimension \(W\) does not participate in the convolution and is
treated as an independent indexing dimension.

Let the input tensor be
\[
X \in \mathbb{R}^{H \times W \times C_{in}}, \quad C_{in} = 1,
\]
and let the convolution filter be
\[
W_f \in \mathbb{R}^{K \times 1},
\]
with bias \(b \in \mathbb{R}\).
Batch and output channel indices are omitted for clarity.

The original convolution output is given by
\begin{equation}
Y(h, w)
=
\sum_{k=0}^{K-1}
W_f(k)\, X(h+k, w) + b,
\label{eq:orig_conv_H}
\end{equation}
where convolution is performed only along the height dimension.

\paragraph{Width Folding Transformation.}
Let \(F\) be the folding factor and assume \(W\) is divisible by \(F\).
We define a transformed input tensor
\[
X' \in \mathbb{R}^{H \times (W/F) \times F}
\]
by re-indexing the width dimension as
\begin{equation}
X'(h, w', f)
=
X(h, Fw' + f),
\quad
f = 0,\ldots,F-1.
\label{eq:width_fold}
\end{equation}
This operation folds the width dimension into the channel dimension
without modifying spatial data along \(H\).

\paragraph{Filter Expansion.}
Since the convolution is independent across width indices, the filter
must be replicated for each folded slice.
We define an expanded filter
\[
W_f' \in \mathbb{R}^{K \times F \times F}
\]
as a diagonal replication:
\begin{equation}
W_f'(k, f, f')
=
\begin{cases}
W_f(k), & f = f', \\
0, & f \neq f'.
\end{cases}
\label{eq:diag_filter_H}
\end{equation}
The bias is similarly replicated as \(b'(f) = b\).

\paragraph{Transformed Convolution.}
The output of the transformed convolution is
\begin{equation}
Y'(h, w', f)
=
\sum_{k=0}^{K-1}
\sum_{f'=0}^{F-1}
W_f'(k,f,f')\, X'(h+k, w', f')
+ b'(f).
\label{eq:new_conv_H}
\end{equation}

Substituting Eq.~\eqref{eq:diag_filter_H} into
Eq.~\eqref{eq:new_conv_H} removes the channel summation:
\begin{equation}
Y'(h, w', f)
=
\sum_{k=0}^{K-1}
W_f(k)\, X'(h+k, w', f) + b.
\label{eq:new_conv_simplified_H}
\end{equation}

Using the folded input definition from
Eq.~\eqref{eq:width_fold}, we obtain
\[
Y'(h, w', f)
=
\sum_{k=0}^{K-1}
W_f(k)\, X(h+k, Fw' + f) + b.
\]

Let \(w = Fw' + f\).
Then
\[
Y'(h, w', f) = Y(h, w),
\]
where \(Y(h,w)\) is the output of the original convolution defined in
Eq.~\eqref{eq:orig_conv_H}.

\paragraph{Conclusion.}
The width folding transformation constitutes a bijective re-indexing of
the width dimension combined with diagonal filter replication.
Since the convolution is performed solely along the height dimension,
the transformation preserves the exact numerical output of the original
network.
Therefore, width folding with channel expansion is
\emph{semantics preserving}.
\hfill\(\square\)

\subsection{N-D Convolutions}
Our method generalizes to N-D convolutions by folding dimensions that are not involved in the convolution operation into the channel dimension. This reparameterization preserves exact convolution semantics via block-diagonal kernels and does not rely on kernel separability or approximation.

\section{Width-Folding Transformation as an IR Transformation} 

\subsection{Motivation}

Formulating width-folding as a compiler transformation is feasible and essential because it decouples the mathematical optimization from any particular library or runtime. By representing the input reorganization and kernel replication at the IR level, the compiler can reason about the data layout, tiling, and vectorization systematically. This enables automatic generation of hardware-efficient code that maximizes utilization of compute units such as Tensor Cores, while preserving the semantics of the original convolution. Treating it as a compiler pass ensures portability, composability with other optimizations, and the ability to target multiple backends without manual intervention.

\subsection{Mathematical Formulation}

Let the input tensor be 
$\mathcal{X} \in \mathbb{R}^{B \times H \times W \times C_{\text{in}}}$, 
and the kernel be 
$\mathcal{K} \in \mathbb{R}^{K_H \times K_W \times C_{\text{in}} \times C_{\text{out}}}$, 
producing output 
$\mathcal{Y} \in \mathbb{R}^{B \times H' \times W' \times C_{\text{out}}}$. 

The \emph{width-folding} transformation reshapes the input tensor along the width dimension.
F (folding factor) is chosen to align with Tensor core tile sizes.

\begin{equation}
\mathcal{X}_f \in \mathbb{R}^{B \times H \times \frac{W}{F} \times (C_{\text{in}} \cdot F)}, \quad
F \in \mathbb{Z}^+ 
\end{equation}

The kernel is correspondingly transformed into a diagonal-blocked form:

\begin{equation}
\mathcal{K}_f \in \mathbb{R}^{K_H \times \frac{K_W}{F} \times (C_{\text{in}} \cdot F) \times (C_{\text{out}} \cdot F)}.
\end{equation}

The transformed convolution preserves the original semantics:

\begin{equation}
\mathcal{Y} = \text{Conv}(\mathcal{X}, \mathcal{K}) = \text{Reconstruct}\big(\text{Conv}(\mathcal{X}_f, \mathcal{K}_f)\big),
\end{equation}

where the reconstruction step involves reshaping the folded output back to the original width dimension.

\subsection{Compiler-Level Realization in MLIR}

Width-folding can be potentially implemented as a {\em semantics-preserving IR transformation in MLIR}. The transformation shall be performed over \texttt{linalg.conv\_2d\_nhwc} or \texttt{linalg.matmul} operations, and potentially consist of the following steps:

\begin{enumerate}
    \item \textbf{Tensor Reshape}: Reshape the input and output tensors to introduce a blocked channel dimension corresponding to the folding factor $F$.
    \item \textbf{Affine Reindexing}: Map the original convolution indices to the folded tensor indices, ensuring that all data dependencies are preserved.
    \item \textbf{Kernel Replication}: Replicate the kernel into a diagonal-blocked layout that is consistent with the folded input tensor.
\end{enumerate}

The legality of this transformation is straightforward: the width dimension must be divisible by $F$, and any padding must be applied consistently to preserve output shape. A lightweight cost model can estimate profitability by considering channel size, tensor core tile alignment, and arithmetic intensity.
The transformation is fully composable with other MLIR passes such as tiling, vectorization, and lowering to CUDA or ROCm backends. Width folding thus provides a mathematically sound, hardware-aware optimization that can significantly improve performance of deep learning kernels in production compilers.

\section{Generalization to GEMM via $1 \times 1$ Convolution}
\label{sec:GEMM}

While width folding is introduced in the context of convolutional operators, the underlying idea extends naturally to general matrix multiplication (GEMM). This follows from the well-known equivalence between GEMM and $1 \times 1$ convolution under appropriate reshaping. We describe this equivalence and show how width folding applies directly to GEMM through this formulation.

\subsection{GEMM as a $1 \times 1$ Convolution}

Consider a matrix multiplication
\[
C = A B,
\]
where $A \in \mathbb{R}^{M \times K}$ and $B \in \mathbb{R}^{K \times N}$.

We reshape matrix $A$ into a tensor
\[
X \in \mathbb{R}^{H \times W \times C_{\text{in}}},
\]
by choosing
\[
H = M, \quad W = 1, \quad C_{\text{in}} = K,
\]
and interpreting each row of $A$ as a spatial position with $K$ channels.

Similarly, matrix $B$ is reshaped into a $1 \times 1$ convolution kernel
\[
W_f \in \mathbb{R}^{1 \times 1 \times C_{\text{in}} \times C_{\text{out}}},
\]
with $C_{\text{out}} = N$.

Applying a $1 \times 1$ convolution produces an output tensor
\[
Y \in \mathbb{R}^{H \times 1 \times C_{\text{out}}},
\]
which is equivalent to the GEMM result $C$ after reshaping.

This construction is algebraically exact and introduces no approximation.

\subsection{Applying Width Folding to GEMM}

In this formulation, the channel dimension corresponds to the reduction dimension $K$ of the matrix multiplication. When $K$ is small or poorly aligned with hardware tile sizes, the resulting computation underutilizes specialized matrix-multiply units.

Width folding can be applied by reindexing an auxiliary spatial dimension and redistributing it into the channel dimension. Concretely, we introduce a synthetic width dimension $W$ and fold it into channels:
\[
X \in \mathbb{R}^{H \times W \times 1}
\;\;\rightarrow\;\;
X' \in \mathbb{R}^{H \times (W/F) \times F}.
\]

The corresponding $1 \times 1$ kernel is replicated into a block-diagonal form, exactly as in the convolutional case. The resulting operator remains a $1 \times 1$ convolution and is therefore equivalent to a GEMM with expanded effective channel dimensionality.

This equivalence shows that width folding is not specific to convolutional layers, but applies more broadly to linear operators expressed as tensor contractions. From a compiler perspective, GEMM and convolution differ only in indexing structure, and both can benefit from operator-level reformulation. This observation further supports expressing width folding as a general compiler transformation rather than a domain-specific optimization.

\section{Potential Optimizations for Sparsity and Quanitization}
Efficient exploitation of sparsity is crucial for both memory savings and computational acceleration. The block-diagonal filter structure introduces structured sparsity that can be leveraged in multiple ways. Only the diagonal blocks of the filters contain non-zero values, while off-diagonal blocks are zero. This structured sparsity allows storing just the non-zero blocks in memory as sparse tensors, significantly reducing memory footprint. For large networks, this can lead to substantial savings, especially in embedded or GPU-constrained environments. Several modern deep learning frameworks and custom CUDA kernels can exploit this by performing sparse matrix multiplications, reducing the number of operations and increasing throughput. Popular deep learning frameworks such as PyTorch and TensorFlow provide built-in support for grouped convolutions. By mapping each diagonal block to a group, the block-diagonal convolution can be implemented efficiently without modifying the core framework. 
    
Structured sparsity pairs naturally with mixed-precision quantization (FP16/INT8). By representing both weights and activations in lower precision, memory bandwidth is reduced, and Tensor Cores can achieve higher throughput. Combining quantization with block-diagonal sparsity ensures that only essential computations are performed at high speed, improving overall efficiency while maintaining model accuracy.


\section{Results}
We implemented the proposed widthfolding transformation in both TensorFlow and TensorRT, validating it against functionally equivalent convolution operations. All experiments and ablation studies were conducted on NVIDIA A100 GPUs using proprietary CNN models. Due to intellectual property and data-sharing constraints, detailed experimental configurations, model architectures, and raw performance measurements cannot be publicly disclosed.
However, a reference implementation in Python using TensorFlow API is provided in Appendix~\ref{app:reference}, allowing reproducibility of the transformation on synthetic or publicly available models. 

To provide context for performance, the complete set of empirical results is summarized in the accompanying patent application ~\cite{roche_pct}. These results demonstrate a minimum of 3× improvement over baseline implementations (vanilla TensorRT C++ API~\cite{tensorrt}) on A100-class hardware. The observed gains are primarily attributable to alignment-aware reformulation of convolution operations, which improves Tensor Core utilization compared to conventional zero-padding strategies.

FP16 precision was used in all experiments. It should be noted that TensorRT does not natively support Tensor Core operations with FP16 input when the number of input channels is not a multiple of eight. It was using a sub-optimal convolution kernel for this case. While H100 and B-series GPUs were unavailable during experimentation, the method is expected to generalize to newer architectures (H200/H800/B200), and re-evaluation on these platforms is planned contingent on hardware access. Importantly, TensorRT is already highly optimized; achieving 3× performance improvement indicates a fundamentally new algorithmic approach, improved hardware utilization, and domain-specific optimization beyond generic kernels.

Although cuDNN ~\cite{chetlur2014cudnn}, another popular deep learning framework, exposes FP16 convolution at the API level, convolutions with very small input channel counts—most notably $C_{\text{in}} = 1$—are not allowed by default due to reason that tensor Cores require matrix dimensions that are multiples of 8. For such cases, the responsibility is pushed to the user rather than the library and users must explicitly pad the channel dimension with zeros or use FP32 variant. Zero padding increases memory traffic and performs wasted computation on artificial data, while falling back to FP32 introduces FP16--FP32--FP16 conversion overhead and executes on CUDA cores (not Tensor cores). In both cases, Tensor Cores remain underutilized, despite their theoretical ability to provide up to an $8\times$ throughput improvement for FP16 operations.

The above facts reflect the design philosophy of TensorRT and cuDNN as a general-purpose kernel library rather than a system that invents new mathematical formulations. They faithfully implement the operation given; they do not reinterpret tensor dimensions or apply semantics-changing transformations to satisfy hardware alignment requirements. In contrast, our approach introduces a semantics-preserving mathematical transformation that increases the effective channel dimension without zero padding or precision up/down-conversion. This reparameterization makes the convolution natively compatible with FP16 Tensor Core kernels, enabling high utilization and substantially higher performance than existing cuDNN or TensorRT implementations for low-channel-count convolutions.

The proposed method applies to any convolutional layer, not just the first layer, and can also be extended to linear layers (i.e., GEMM) as noted in Section~\ref{sec:GEMM}. This extension implies that the transformation could accelerate GEMM operations in cuBLAS (especially, those involving {\em tall skinny matrices}). A $3\times$ speedup over cuBLAS is a potential break through in compiler optimization.

The observed results are consistent with publicly available NVIDIA documentation~\cite{nvidia_cnn} , providing further confidence in the correctness and potential impact of the method. Finally, this transformation can be applied even during tranining process to realize similar speedups and tensorcore utilization.

\section{Related Work}
Optimizing CNNs for hardware accelerators has been an active area of research. Our approach draws inspiration from several exising techniques, combining channel expansion and block-diagonal sparsity to fully exploit hardware throughput. 

\subsection{Relation to Other CNN Variants}

\subsubsection{Grouped Convolutions}
In grouped convolutions~\cite{krizhevsky2012imagenet}, input channels are divided into separate groups, and each group is convolved independently with a corresponding set of filters. Our block-diagonal transformation naturally induces a similar structure: each diagonal block in the filter corresponds to a “group” that processes a subset of the input channels. However, unlike standard grouped convolutions, our approach preserves the original filter values within each block and replicates them in a controlled manner to align with Tensor Core hardware. This ensures maximum hardware utilization without changing the semantic meaning of the original convolution.

\subsubsection{Depthwise and Pointwise Convolutions}
Depthwise convolutions~\cite{howard2017mobilenets} apply a single filter per input channel, significantly reducing computation but also limiting feature interactions across channels. Pointwise (1×1) convolutions are then used to combine features across channels. The block-diagonal filter transformation resembles a hybrid of these approaches: each block is multi-channel (not single-channel as in depthwise) and operates on a subset of input channels, while off-diagonal blocks are zeroed to create structured sparsity. This design maintains full multi-channel processing within blocks while still enabling computational efficiency similar to depthwise convolutions. Unlike traditional depthwise or pointwise convolutions, our method explicitly targets hardware alignment for Tensor Cores.

\subsubsection{Channel Expansion / 1×1 Convolutions}
Channel expansion via 1×1 convolutions is commonly used to increase the number of output channels and enable more expressive representations. In our approach, the block-diagonal transformation can be interpreted as a structured channel expansion: the original output channels are replicated across diagonal blocks to create a padded, hardware-aligned output tensor. This provides the benefits of channel expansion—higher representational capacity—without introducing additional learnable parameters beyond the original filter. The key distinction is that our expansion is carefully organized to match hardware constraints, which is not considered in conventional 1×1 convolutions.

Similar strategies have been explored in low-rank filter approximations~\cite{jaderberg2014speeding} and kernel tiling optimizations~\cite{chetlur2014cudnn}, but our method maintains the original filter semantics while providing full Tensor Core alignment.

\subsection{Difference from Channel Zero-Padding}

A simple approach to align channels for hardware is \emph{zero-padding}: if a layer has 1 input channel and the hardware prefers multiples of 8, 7 zero channels are added. While this aligns the total channels, these extra channels carry no information, wasting compute on irrelevant data. In contrast, \emph{width-folding} avoids adding input channels, expanding only the (small) weight tensor. It creates \emph{block-diagonal filters}, introducing structured sparsity that frameworks can exploit for efficient computation. Moreover, this pattern naturally generalizes to higher-dimensional and grouped convolutions.

\subsection{Contrast with Traditional Compiler Transformations}

Width folding differs fundamentally from traditional compiler
transformations applied in deep learning systems.
Traditional compiler optimizations—such as operator fusion, loop tiling,
layout reordering, vectorization, and kernel selection—preserve the
\emph{high-level mathematical definition} of an operator and optimize
\emph{how} that computation is executed on hardware.
These transformations act on scheduling, memory access patterns, or code
generation, but do not alter the underlying convolution formulation.

In contrast, width folding operates at the level of the \emph{mathematical
operator itself}.
The transformation explicitly rewrites the convolution into an equivalent
form by re-indexing tensor dimensions and restructuring filter parameters.
This reformulation changes the apparent tensor shapes and filter structure
seen by the compiler, while preserving exact input--output semantics by
construction.
As a result, hardware alignment constraints (e.g., channel dimensions being
multiples of a fixed factor) are satisfied without relying on padding or
special-case kernel implementations.

Finally, while compiler transformations are generally opaque and
hardware-specific, width folding is expressed as an explicit, semantics-
preserving rewrite rule.
This makes the transformation analyzable, provably correct, and amenable to
future automation.

Width folding is a \emph{post-training, pre-compilation} transformation:
it modifies the trained model itself before it is handed to the compiler or
runtime.  This enables downstream compilers to treat the transformed model as a
standard convolution that naturally maps to optimized hardware kernels.
Rather than replacing compiler optimizations, width folding complements
them by expanding the space of hardware-friendly formulations that
compilers can further optimize.

\section{Conclusion}
We presented a transformation for CNN filters that aligns input and output channels with NVIDIA Tensor Core requirements. The method maintains the original filter structure, generalizes to n-D convolutions, and exploits sparsity for memory and computational efficiency. This approach bridges hardware-aware optimizations with traditional CNN design principles, providing a practical and scalable solution for high-performance deep learning deployments.

\subsection{Future Work}

Several promising research directions follow naturally.
First, we plan to explore a broader class of \emph{structure-preserving
transformations} beyond block-diagonal reformulations. These include
alternative spatial--channel folding strategies, hierarchical blocking,
mixed-radix reshaping schemes, sparsity aware folding and dilation
rewriting. Such transformations could enable
efficient utilization of hardware units with strict alignment,
vectorization, or tiling requirements. Also, it is interesting
to study if the inverse (i.e. channel-to-space) is possible mathematically.

Second, we plan to investigate
transformations tailored to other GPU architectures, including AMD
GPUs, where wavefront sizes, memory coalescing rules, and matrix
instruction formats differ substantially. Developing a unified
transformation framework that adapts this idea to multiple
specialized hardware as well as general-purpose CPU (e.g. alignment
requirement for SIMD or cache usage ~\cite{onednn2025}) 
remains an open and important problem.

Third, we aim to extend these techniques to other operators commonly
used in modern architectures, such as grouped convolutions,
depthwise separable convolutions,
attention mechanisms~\cite{attention}, and recurrent layers~\cite{lstm}. 

Finally, we envision integrating these transformations into MLIR~\cite{mlir} framework. More broadly, we introduce \emph{semantic tuning} as a post-training optimization paradigm in which the mathematical formulation of a neural network is rewritten to better match hardware execution constraints while preserving the exact input–output semantics of the original model. This shifts the compiler optimizations from hardware-specific kernel
optimization toward \emph{semantically equivalent operator transformations}, 
achieving performance portability.

\FloatBarrier
\bibliography{cnn_paper_icml}
\bibliographystyle{icml2026}

\newpage
\appendix
\onecolumn
\section{Example Code using Tensorflow CNN (Channels-Last)}
\label{app:reference}
This Tensorflow implementation demonstrates a 2-D block-diagonal filter transformation where the input is in channels-last format ($[B, H, W, C_in]$), the W dimension is folded, and the convolution is compatible with NVIDIA Tensor Cores.

\lstset{
    language=Python,
    basicstyle=\ttfamily\small,
    keywordstyle=\color{blue},
    commentstyle=\color{green!60!black},
    stringstyle=\color{orange},
    numbers=left,
    numberstyle=\tiny\color{gray},
    stepnumber=1,
    numbersep=5pt,
    showspaces=false,
    showstringspaces=false,
    breaklines=true,
    frame=single,
    caption={Width-folding CNN transformation in TensorFlow (channels-last).},
    label={lst:width_folding}
}

\begin{lstlisting}
import tensorflow as tf
import numpy as np

# -----------------------------
# Parameters
# -----------------------------
B, H, W = 1, 32, 64     # batch, height, width
K = 5                  # kernel size along H
F = 8                  # width folding factor
Cout = 1               # output channels

assert W % F == 0

# -----------------------------
# Input tensor (NHWC)
# -----------------------------
x = tf.random.normal((B, H, W, 1))

# -----------------------------
# Original filter + bias
# Conv along H only -> kernel (K,1)
# -----------------------------
filterVal = tf.random.normal((K, 1, 1, Cout))
biasVal = tf.random.normal((Cout,))

# -----------------------------
# Original convolution
# -----------------------------
y_orig = tf.nn.conv2d(
    x,
    filterVal,
    strides=[1, 1, 1, 1],
    padding="VALID",
    data_format="NHWC"
)
y_orig = tf.nn.bias_add(y_orig, biasVal)

# -----------------------------
# Width folding: W -> W/F, Cin -> F
# -----------------------------
# (B, H, W, 1) -> (B, H, W/F, F)
x_folded = tf.reshape(x, (B, H, W // F, F))

# -----------------------------
# Build diagonal filter
# (K,1,1,Cout) -> (K,1,F,F*Cout)
# -----------------------------
filterValNew = np.zeros((K, 1, F, F * Cout), dtype=np.float32)

for f in range(F):
    filterValNew[:, :, f, f*Cout:(f+1)*Cout] = np.squeeze(filterVal.numpy(),axis=-1)

filterValNew = tf.constant(filterValNew)

# -----------------------------
# Bias replication
# -----------------------------
biasValNew = tf.tile(biasVal, [F])

# -----------------------------
# Folded convolution
# -----------------------------
y_folded = tf.nn.conv2d(
    x_folded,
    filterValNew,
    strides=[1, 1, 1, 1],   # stride only along H
    padding="VALID",
    data_format="NHWC"
)
y_folded = tf.nn.bias_add(y_folded, biasValNew)

# -----------------------------
# Reconstruct original layout
# (B, H', W/F, F) -> (B, H', W)
# -----------------------------
y_reconstructed = tf.reshape(
    y_folded,
    (B, y_folded.shape[1], W)
)

# -----------------------------
# Verification
# -----------------------------
max_error = tf.reduce_max(tf.abs(np.squeeze(y_orig, axis=-1) - y_reconstructed))
print("Max absolute error:", max_error.numpy())

# Assert correctness
tf.debugging.assert_near(np.squeeze(y_orig, axis=-1), y_reconstructed, atol=1e-5)
print("Width folding transformation is numerically correct")
\end{lstlisting}

\end{document}